\documentclass[10pt]{wlscirep}
\usepackage[utf8]{inputenc}
\usepackage[T1]{fontenc}
\usepackage{xcolor}
\usepackage[version=3]{mhchem}
\usepackage{multirow}
\usepackage{float}
\usepackage{soul}
\usepackage{fontawesome}
\usepackage{array}
\usepackage{gensymb}
\usepackage{upgreek}
\usepackage{xfrac}
\usepackage{lscape}

\newcommand{\beginsupplement}{%
        \setcounter{table}{0}
        \renewcommand{\thetable}{S\arabic{table}}%
        \setcounter{figure}{0}
        \renewcommand{\thefigure}{S\arabic{figure}}%
     }

\usepackage{newfloat}
\DeclareFloatingEnvironment[name={Supplementary Figure}]{suppfigure}

\title{Sub-Pixel Electron Beam Alignment for Machine Learning Characterization of Hybrid Pixel Detectors}

\author[1,*]{Emiliya Poghosyan}
\author[2]{Xiangyu Xie}
\author[3]{Joakim Reuteler}
\author[2]{Kirsty A. Paton}
\author[4]{Luis Barba Flores}
\author[4]{Benjamin B\'ejar Haro}
\author[2]{Erik Fröjdh}
\author[2]{Anna Bergamaschi}
\author[1]{Elisabeth Müller}

\affil[1]{PSI Center for Life Sciences, Paul Scherrer Institute, Villigen PSI, 5232, Switzerland}
\affil[2]{PSI Center for Photon Science, Paul Scherrer Institute, Villigen PSI, 5232, Switzerland}
\affil[3]{ScopeM, ETH Zürich, Otto-Stern-Weg 3, 8093, Switzerland}
\affil[4]{Swiss Data Science Center, Paul Scherrer Institute, Villigen PSI, 5232, Switzerland}

\affil[*]{emiliya.poghosyan@psi.ch}


\begin{abstract}
Due to their radiation hardness, kilohertz frame rates, and high dynamic range, hybrid pixel detectors have recently expanded their application range to electron diffraction and recently also electron imaging. However, these detectors typically have pixel sizes about ten times larger than those of direct electron detectors commonly used for imaging and more prominent electron multiple scattering effects. To overcome these limitations, machine learning approaches can be utilized to reconstruct the electron entrance point and achieve super-resolution. As this process is inherently stochastic, and machine learning relies on suitable training data, high-quality, representative training data are essential for developing models that achieve the best possible resolution.
In this work, we present two novel experimental methods for generating such training data. The first method employs precise microscope alignment to scan the detector plane using a finely focused electron beam of 2 $\mu$m diameter, enabling controlled sub-pixel mapping. The second method utilizes specially designed aperture masks with sub-pixel-sized holes to accurately localize electron entry points. We developed and validated two experimental strategies for collecting training data at acceleration voltages of 60, 80, 120, and 200 keV, which enable sub-pixel labeling for hybrid pixel detectors. 
Notably, our methodology is broadly applicable to a wide range of hybrid pixel detectors.
\end{abstract}
\begin{document}

\flushbottom
\maketitle

\thispagestyle{empty}
\section{Introduction}
The development and application of hybrid pixel detectors in electron microscopy have substantially expanded the capabilities and scope of electron microscopy across different research fields. These include the use of fast pixelated detectors in 4D-STEM and electron ptychography\cite{Pennycook2015, Pelz2017, Zhou2020, Chen2021, Sha2023, Pelz2023SolvingTomography, Skoupy2025Ptychoscopy:Ptychography}, applications in electron diffraction\cite{Gruene2018SchnelleElektronenbeugung, Mahmoudi2025ExperimentalDiffraction}, and, in some studies, extensions into imaging modes\cite{vanSchayck2023IntegrationWorkflow, McMullan2023}. Recent experimental and theoretical studies in single-particle cryo-electron microscopy have demonstrated that an operating voltage of 100 keV is optimal, providing the best balance between information transfer and radiation damage compared to the higher acceleration voltages traditionally employed in the field\cite{McMullan2023}. While advances in widely used Monolithic Active Pixel Sensors  have expanded their performance at lower voltages\cite{Chan2024High-resolutionDetector}, their frame rate, radiation hardness, and near-ideal detector quantum efficiency (DQE) still do not match those of hybrid pixel detectors. Therefore, the prospect of a radiation-hard, universal detector capable of seamlessly switching between real and reciprocal space is highly appealing, but its realization depends critically on improving the spatial resolution of hybrid pixel detectors.

The primary limitation arises from the pixel size—typically greater than 50 $\mu$m in commercially available devices. 
At higher electron energies, such as 200–300 keV commonly used in electron microscopy, the incident electrons do not deposit their energy within a single pixel but instead produce charge tracks spanning multiple pixels, thereby degrading spatial resolution. 
Accurately determining the electron impact position within a single pixel, i.e. in the sub-pixel regime, is therefore essential for resolution enhancement. 
However, the stochastic nature of electron energy deposition toward the end of its trajectory makes it challenging to apply conventional center-of-mass approaches to retrieve the impact position. Some convolutional neural network based approaches\cite{vanSchayck2023IntegrationWorkflow} have been proposed to address this challenge; however, to our knowledge, all have relied solely on synthetic training data, leading to suboptimal performance and limited transferability due to the mismatch between synthetic and real data distribution. Therefore, using experimental training data acquired on the respective detector enables including the respective detector properties.

The aim of this work has been to advance the spatial resolution capabilities of hybrid pixel detectors by developing and detailing sub-pixel electron beam alignment strategies, in order to enable direct, reproducible acquisition of detector specific experimental training data that capture physical effects and imperfections beyond the reach of simulation, thereby establishing a foundation for robust and transferable machine-learning calibration of hybrid pixel detectors. We have applied these strategies to a prototype hybrid pixel detector, Mönch\cite{Dinapoli2014}, featuring a 25 $\mu$m pixel pitch. Our previous study\cite{Xie2024EnhancingLearning} demonstrated the feasibility of machine-learning–based sub-pixel reconstruction using both synthetic and experimental training data acquired at 200 keV.  With this work, we provide, for the first time, a comprehensive methodological framework for generating such training data describing the procedures used to obtain the experimental datasets and extending these approaches to multiple acceleration voltages (60, 80, 120, and 200 keV).

\section{Results}
Experimental data with precise labeling are essential for effective machine learning (ML) model training, as even state-of-the-art simulations fail to replicate the full variability of detector response, and therefore, potentially, limiting the model's precision in a real setting.

\begin{figure*}[htbp]
\centering
\includegraphics[width=0.8\textwidth]{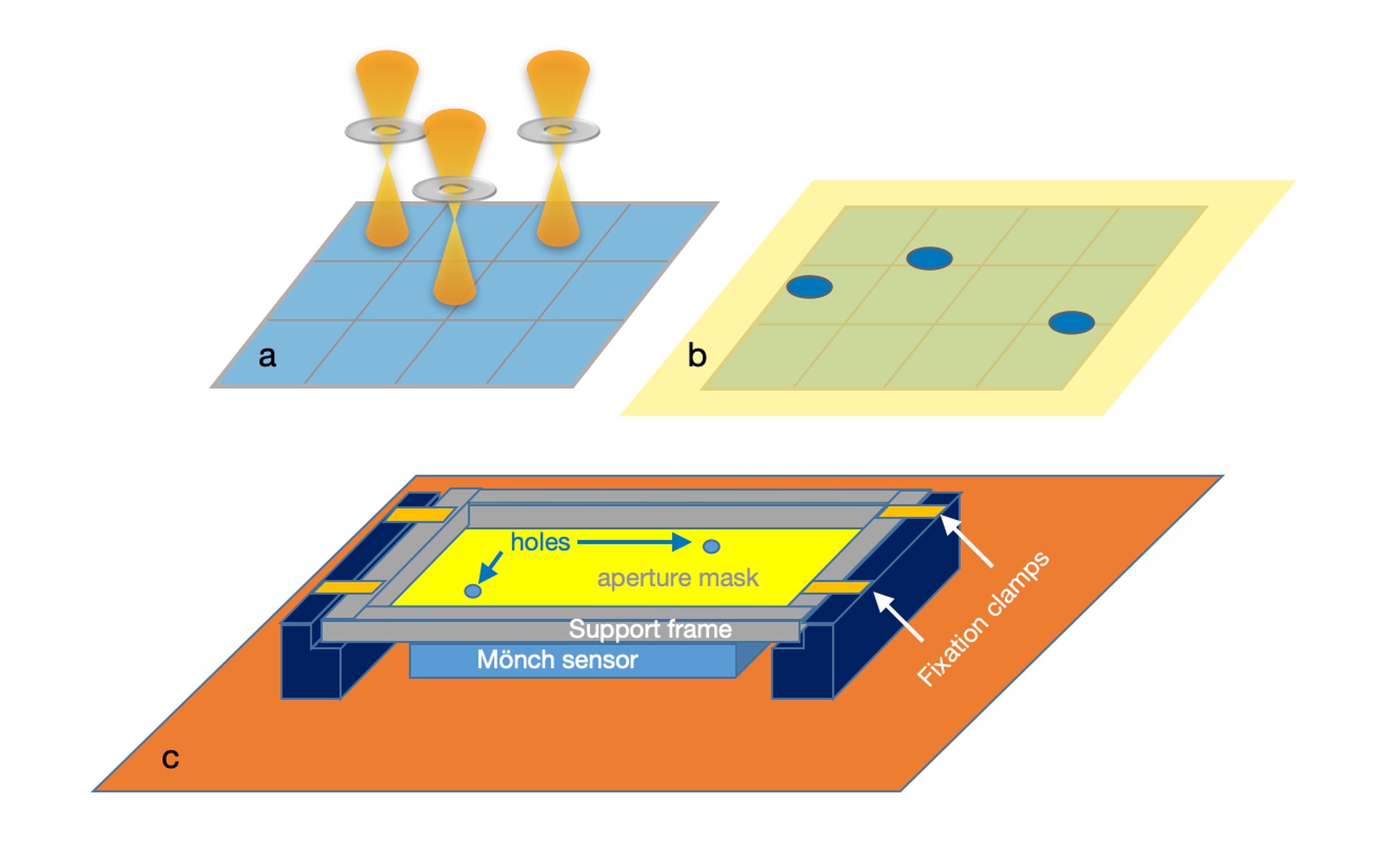}
\caption{\textbf{Schematic representation of the detector characterization strategies} projection of the real space image of a narrow beam onto the detector (a–b) and the aperture mask setup on the MÖNCH sensor (c). For measurements at 200 keV, a narrow, semi-parallel beam with a diameter around 2 $\mu$m was used (a). For a more universal method, an aperture mask with holes of varying diameters between 2 and 5 $\mu$m was designed (b). The mask can be mounted above the sensor using fixation clamps on a support frame, as illustrated in (c).}
\label{fig:scheme}
\end{figure*}

Accordingly, we developed two independent methods for generating experimental training data for detector calibration. The first method utilizes a narrow, low-intensity electron beam diameter ($\simeq$2 $\mu$m) to systematically probe the effect of the electron impact position within a single pixel at 200 keV (Figure \ref{fig:scheme}a). The second method, broadly applicable across acceleration voltages, employs a masked aperture (Figures \ref{fig:mask1} and \ref{fig:masks}) mounted directly above the detector chip (Figure \ref{fig:scheme}c), physically restricting the impact positions. This enables the experimental determination of electron impact positions within 2–5 $\mu$m, as set by the mask geometry. This method was implemented for detector characterization at 60, 80, 120, and 200 keV (Figure \ref{fig:scheme}b). Both experimental datasets were compared with simulated training data, allowing for direct comparison between the two approaches performed at 200 keV.

\subsection{\textbf{Focused beam based detector calibration at 200 keV}}

To generate a narrow focused beam on the detector plane, we developed a custom microscope alignment on our probe-corrected JEOL NeoARM 200F instrument. The idea being similar to confocal geometry that has been previously used for mapping the response of annular dark field (ADF) detectors\cite{LeBeau2008ExperimentalMicroscopy, Jones2018AnPre-field, JonesPracticalRecording}.
This alignment was configured to project the image plane of the objective lens, rather than its back focal plane, i.e. a Ronchigram, onto the detector. Starting with a nearly parallel beam featuring a very low convergence semi-angle (1–3 mrad), we systematically adjusted the intermediate and projector lens system to achieve the narrowest possible beam on the detector. Using the Free Lens Control functionality, we first switched off the Projector Lens (PL) and adjusted Intermediate Lens 3 (IL3) to minimize the beam diameter on the detector plane. This alignment produced a signal primarily from beam absorption through the sample. To bring the sample image into focus, we subsequently fine-tuned the Condenser Lens 3 (CL3) current.

For beam diameter calibration, the STEM scan of a standard cross-grating with latex beads (Ted Pella Inc. 2,000 lines/mm, product number 673) was projected onto the detector, enabling the calibration. However, the initial alignment resulted in a very small scanning area, making it difficult to clearly resolve individual grid squares required for accurate calibration. To increase the scanning range, we adjusted the current through Intermediate Lens 2 (IL2), which slightly broadened the beam diameter. This increase was compensated by further tuning PL and IL3. Using this procedure, we measured the magnification between the sample and detector plane. Therefore, while knowing the beam diameter on the sample plane as well as the magnification applied, we could calculate the beam diameter at the detector plane to be approximately 2 $\mu$m.

To scan the aligned beam across the sensor, a custom Python script utilizing the PyJEM API was implemented to control the shift coils of the scanning unit and send data acquisition signals to the detector. A dwell time was set to acquire 1,000 frames per scan position. Assuming that all tracks of the incident electrons started from the same point but propagated through the silicon sensor undergoing statistical angular deflections due to multiple scattering, the average of charge centroid positions for the different tracks determined the entry points of the corresponding electron events, providing an estimate with negligible bias and uncertainty. Having the ground truth of the entry point position, allowed us to apply supervised learning with neural networks to estimate the true entry point for each track.

By randomly sampling the beam position more than 400 times (see Figure \ref{fig:scan_mask}a), we ensured the generalization power of the training dataset by a uniform and dense distribution of sub-pixel beam positions and the use of data augmentation for all available symmetries to further enhance the generalization power.

\subsection{\textbf{Detector calibration with broad beam illumination}}
To generate a training dataset independent of complex microscope alignments and transferable across different instruments and acceleration voltages, a universal aperture mask was designed for detector characterization (Figures \ref{fig:mask1} and \ref{fig:masks}). This mask included perforated holes of various geometries, as well as a knife edge along one side for modulation transfer function (MTF) and detective quantum efficiency (DQE) measurements.

Prior to fabrication, Monte Carlo simulations (Figure \ref{fig:shielding}) using the Allpix Squared framework\cite{Spannagel2018Allpix2:Detectors} were performed to determine the optimal mask material and thickness for acceleration voltages between 60 - 200 keV, ensuring electron transmission only through the intended holes. Tungsten was selected as the mask material, with a thickness of either 20 $\mu$m or 25 $\mu$m depending on the specific mask described below.

\subsubsection{\textbf{Mask micro-fabrication using FIB/SEM}}

\begin{figure*}[htbp]
\centering
\includegraphics[width=0.8\textwidth]{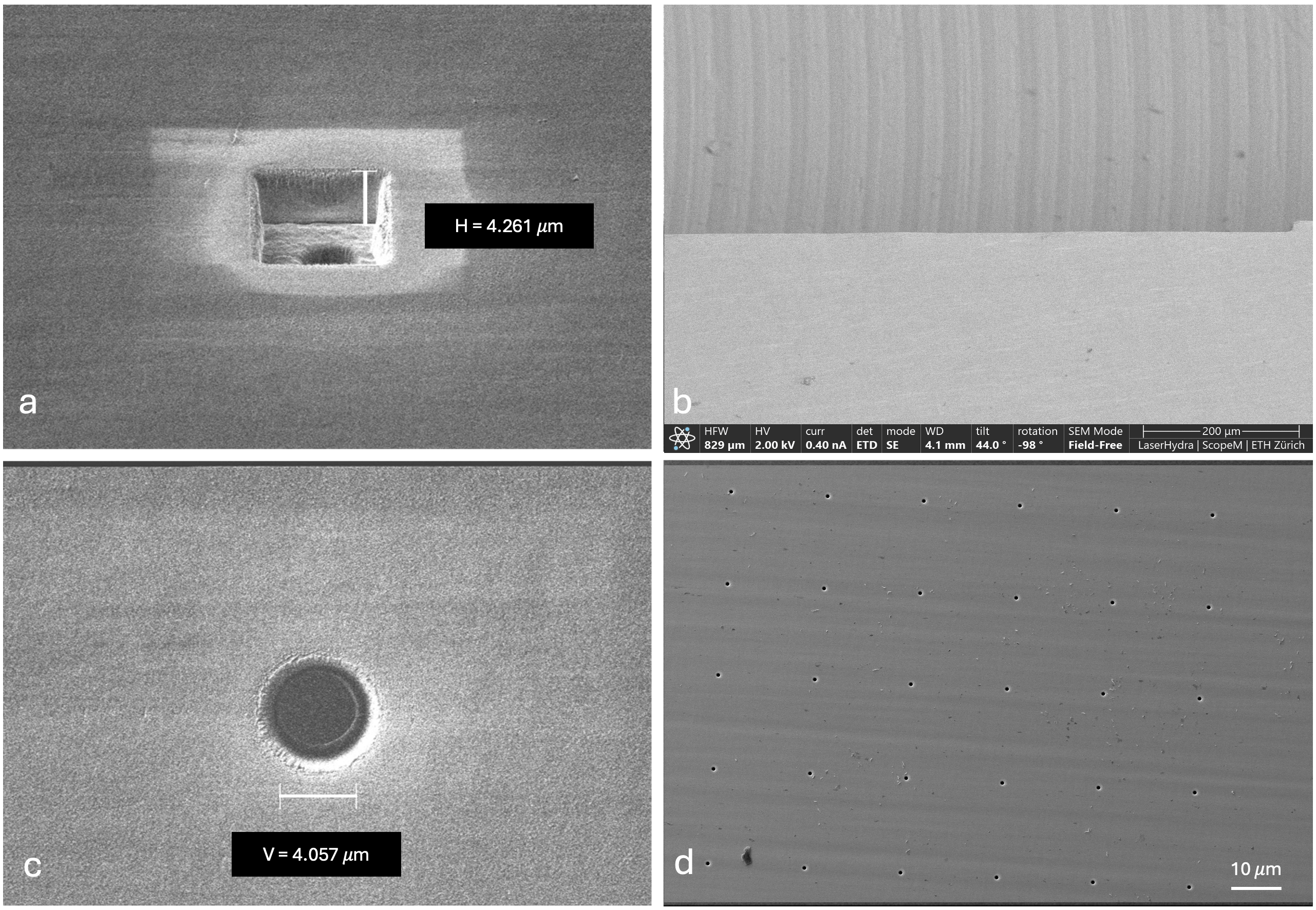}
\caption{\textbf{Manufacturing of holes (a, c, and d) and the knife edge (b) was performed using Ga and plasma FIB/SEM respectively.} Two types of holes were tested. One hole geometry shown in (a) featuring a groove with a height of approximately 4 $\mu$m. The other type were just straight holes, penetrating the whole W-foil with as little change in diameter as possible (c). The holes are arranged as a regular 2D array (d).}
\label{fig:mask1}
\end{figure*}

We iterated the mask design as detailed in the following section, completing three design iterations. The initial mask design, seen in Figure \ref{fig:masks}a, featured small holes, 3–5 $\mu$m in diameter, patterned using a focused ion beam instrument (Zeiss Nvision40 Galium FIB/SEM) Figure \ref{fig:mask1}, and spaced sufficiently apart to ensure that electron trajectories from different holes remained separated (see Table \ref{tab1}). The distance between holes was chosen to be large enough to prevent overlap between electron tracks from adjacent holes and the distance needed to slightly exceed an integer multiple of the sensor pixel pitch to distribute their sub‑pixel positions uniformly. The aperture masks were initially attached to an aluminum frame using silver glue, which was then mounted less than 1 mm above the sensor (see Figure \ref{fig:scheme}c). The design facilitated mask exchange within approximately one hour, minimizing microscope downtime. However, as the fabrication time per hole using gallium FIB/SEM was approximately fifteen minutes, we employed a plasma FIB/SEM (Thermo Fischer Scientific Helios 5 Laser Hydra) for subsequent masks which allowed for significantly increased hole numbers. The tungsten foils were mounted on silicon frames prior to mask milling because the aluminum frames were too large for some holes at the edges, causing the stage to touch SEM lens at the required large tilt angles. The Si-frame was then fixed with silver glue on the Al-frames with the W-foil being at the bottom side of the Si-frame, i.e. closest possible to the aluminum frame.
The holes were milled with 30 kV, 60 nA Xenon ion beam using a 5 $\mu$m diameter disk pattern, taking approximately one minute per hole. The openings measured on the backside of the W-foil were about 4.5 $\mu$m in diameter.

Due to the very large number of holes needed (see Number of holes in Table \ref{tab1}), an automated approach for hole fabrication was chosen. Mask fabrication relied on a custom Python script to control stage positioning and milling pattern. In order to minimize the movement of the stage when transitioning from one hole to the next a serpentine path was chosen. The code allowed setting the desired pitch and number of rows and columns for a rectangular hole array. For Mask 3, two such rectangular arrays were used in sequence, in order to create the L-shaped hole array without interruption causing an unwanted discrepancy in the hole positions.
Since the thin W-foils were not completely flat, the stage Z position was fixed at a medium level fitting sufficiently well for the whole foil area to be processed. Once the patterning by the script was done, the resulting hole arrays were imaged at high pixel density to provide a precise chart of the hole positions (see Figure \ref{fig:mask1}d).

Prior to the data acquisition using the broad beam approach, the electron flux was calibrated with the help of a Faraday cup, attached to the small fluorescent screen in NeoARM 200F microscope. The flux was reduced so that, on average, only one electron or none reached the sensor per hole per image frame. This translated into an occupancy of a few percent per pixel, where we relied on the acquisition of a large number of frames to ensure sufficient statistics. Since the holes were much smaller than the 25 $\mu$m sensor pixels, and the electrons scattered randomly in the sensor, summing a full dataset from repeated images as well as a prior knowledge of the holes position relative to each other enabled unambiguous identification of each pixel where electrons entered the sensor. This was possible due to the design of the edges of the masks (e.g. (x,y) and (x$_1$,y$_1$) coordinates in Figure \ref{fig:masks}a), with pre-measured distances in the SEM relative to the holes and the knife edge.

\subsubsection{\textbf{Iterative improvements of the aperture mask design}}

\begin{figure*}[htbp]
\centering
\includegraphics[width=0.8\textwidth]{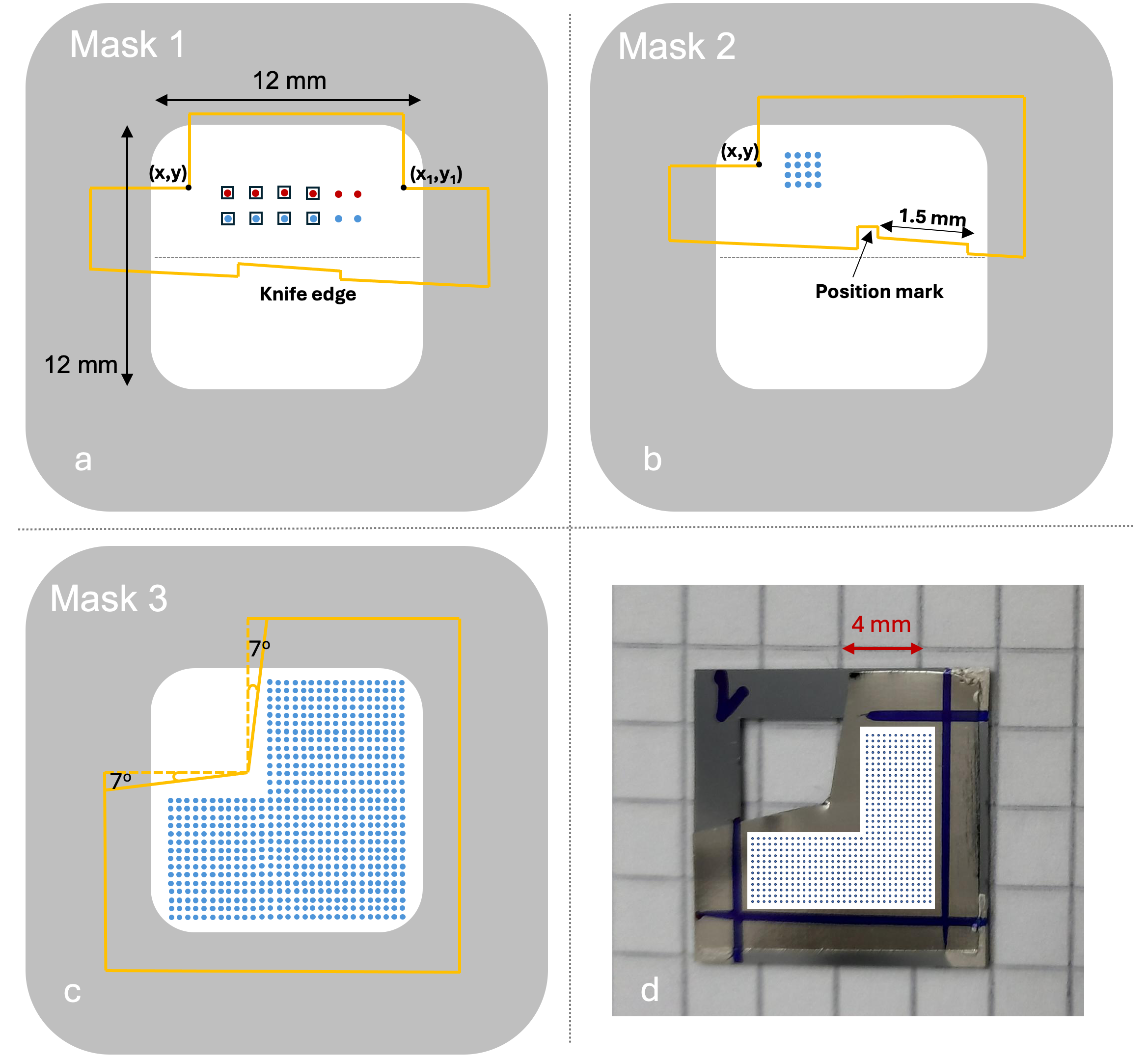}
\caption{\textbf{Graphic representation of mask aperture evolution (a – c) and a photograph of the final mask (d) with an overlay positions of the holes, mounted on a silicon frame support for FIB/SEM.} Mask 1 featured two corners of different sizes (a) to aid in determining the distance to the holes after mounting above the detector. Mask 2 had a single open corner and an additional position marker next to the knife edge (b), along with a greater number of holes. Note that the diagrams show  rather the positioning than actual number of holes. For mask 3 the W-foil covered a larger surface area and contained an even higher number of hole  (c). All masks were mounted on an aluminum plate with an opening of approximately 12 × 12 mm; Masks 2 and 3 were first glued to a silicon support. The foil shape is outlined in yellow, with holes marked as red circles (3 $\mu$m) or blue (5 $\mu$m). Blue rectangles around certain holes in Mask 1 indicate additional wells milled around the respective holes.}
\label{fig:masks}
\end{figure*}

\begin{figure}[!htbp]
\centering
\includegraphics[width=0.35\textwidth]{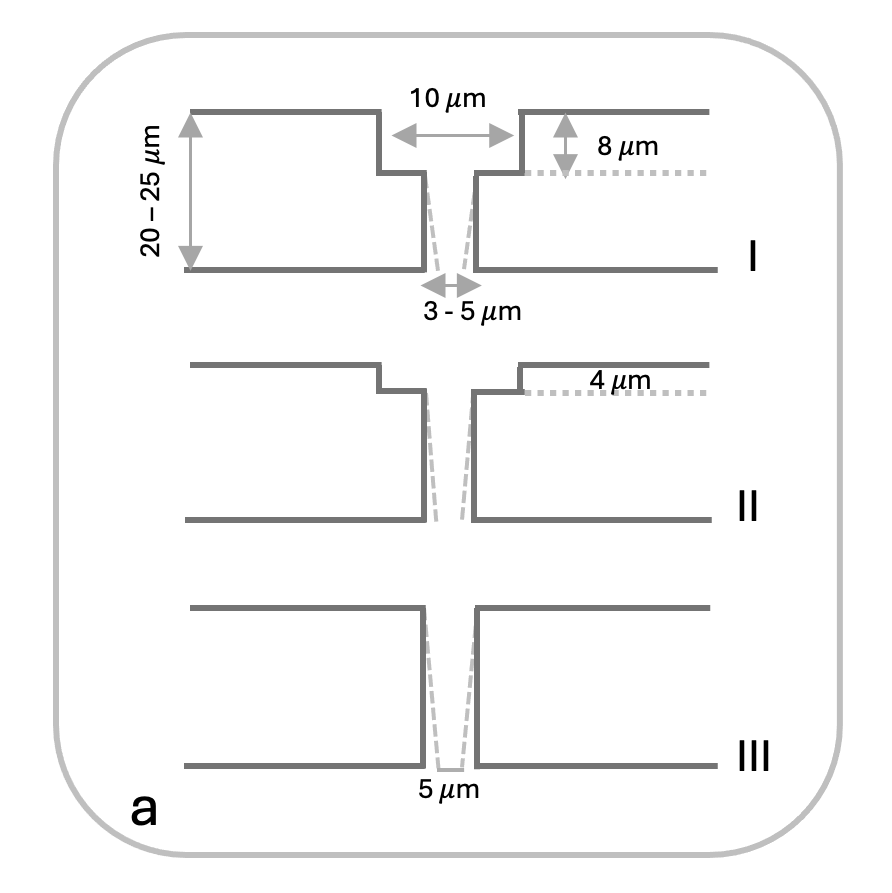}
\caption{\textbf{Schematic diagram of the hole geometries used in this work.} Panels I, II and III show the holes included in the first mask geometry, where wells of different depths were first milled before the actual hole was cut into the W-foil. Hole geometry III was used for masks two and three, since the holes in the first mask had confirmed that despite the large thickness of the W-foil it was possible to drill a hole through the full thickness of the membrane with the FIB. For simplicity, the drawing represents the hole as a cylinder, whereas in reality, due to the beam shape, it more closely resembles a trapezoid with some narrowing at the bottom, as indicated with a dashed lines.}
\label{fig:holes}
\end{figure}

Our initial aperture mask design was largely exploratory. We used a 20 $\mu$m thick tungsten foil and fabricated a total of 12 holes with varying geometries (Figure \ref{fig:mask1}). The hole diameters ranged from 3 to 5 $\mu$m, with a fixed spacing between them. Additionally, wells of 10 $\mu$m diameter and varying depths (4 and 8 $\mu$m) were prepared to evaluate electron signal penetration at different acceleration voltages (see Figure \ref{fig:holes}), as well as to ensure an appropriate aspect ratio between width and length for fabricating holes of such small diameter. The knife edge length was approximately 1 mm. Gallium or xenon ion milling was performed perpendicular to the mask surface in case for the holes, as well as for the knife edge, in order to ensure the cross section was as flat/abrupt as possible. 

In addition, the mask aperture plate covered only about half of the detector, allowing the remaining area to be used for calibration procedures, such as acquiring the knife-edge profiling and noise power spectra for MTF and DQE analysis. Notably, some electrons seemed to bleed-through  the mask at 200 keV when at the knife-edge half of the beam was incident on the mask (without holes), while the other half was transmitted directly onto the detector. One possible explanation is that electrons undergo multiple scattering at the mask edge and are finally back scattered into the detector.

To minimize the possibility of bleed-through caused by mask thickness, the second and third mask design employed a 25 $\mu$m thick tungsten foil. For statistical robustness and training data sampling, having only 12 holes was insufficient. Consequently, the second design featured a larger aperture mask surface area with an increased number of holes (see Table \ref{tab1}). To achieve this, the mask shape was redesigned while maintaining sufficient distance between the knife edge and the initial holes. Only one corner of the foil was left uncut to define the precise location of the holes within the mask (Figure \ref{fig:masks}b). Additionally, the second design incorporated holes of fixed diameter and omitted the wells from the first mask.
This was because, despite the high aspect ratio, the size gradient of the holes from top to bottom was sufficiently small to be acceptable.
Moreover, the reduced foil thickness in the region of the pre‑milled well used in the first mask design, increased the risk of incomplete blocking of incident electrons from reaching the sensor.

As expected, the second mask design significantly improved the statistical quality of the training dataset; however, electron track crossing was observed due to still insufficient hole spacing, which was much more prominent with larger statistics. This was the case for acceleration voltages of 120 and 200 keV. Therefore, to address this, we developed a third and final mask design, further increasing the mask surface area and the number of holes while simultaneously enlarging the distance between neighboring holes to minimize track overlaps (see Table \ref{tab1} and Figure \ref{fig:masks}c, d).

\subsection{\textbf{Two experimental training data acquisition strategies at 200 keV}}\label{subsec1}

\begin{figure}[htbp]
\centering
\includegraphics[width=0.5\textwidth]{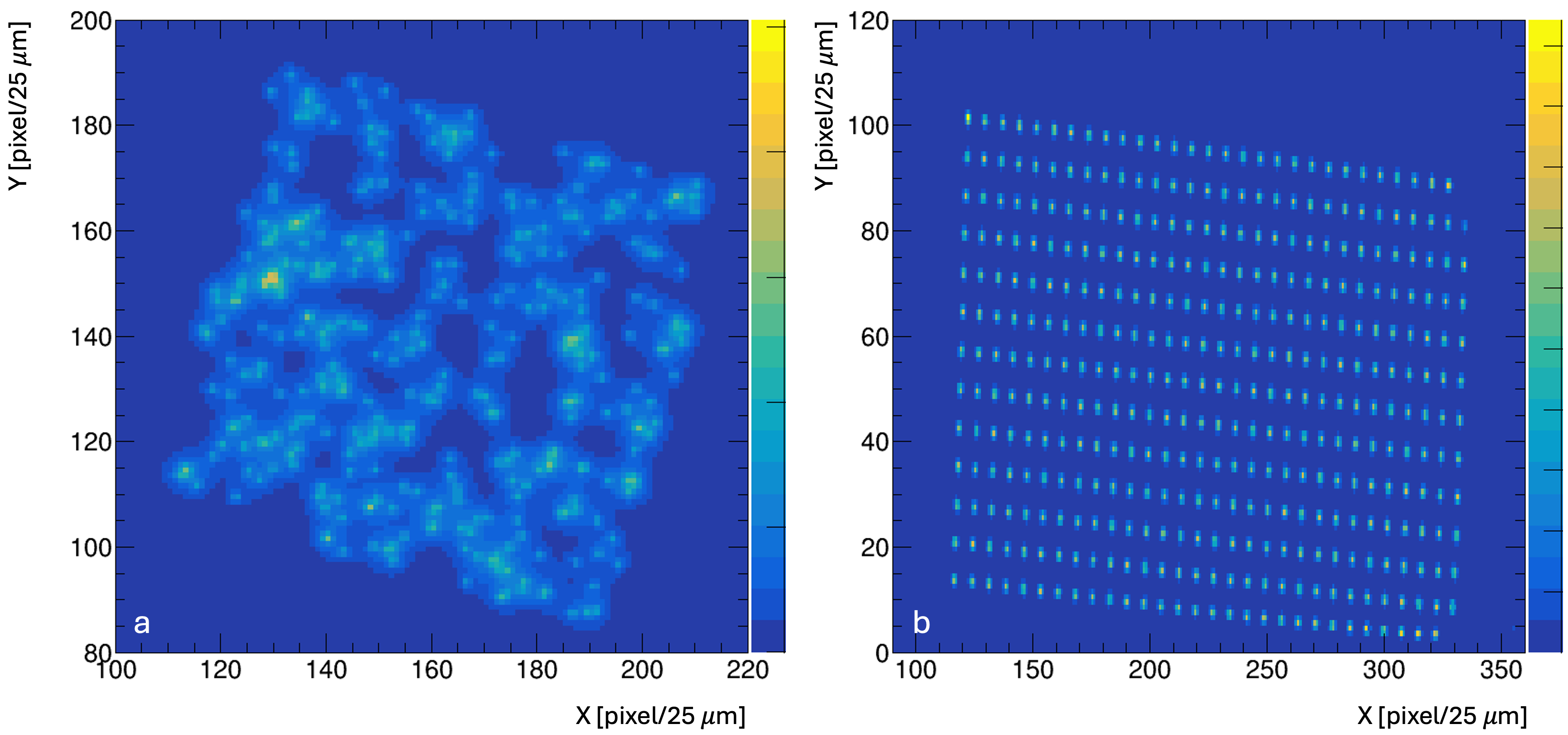}
\caption{\textbf{Example of detector signal at 200 and 80 keV obtained from two different training data acquisition strategies.} (a) Overlay of scanning points acquired using focused beam-based detector calibration at 200 keV, where each scanning point represents thousand of electron events. (b) Aperture Mask 2 array under parallel beam illumination at 80 keV, used to fit grid coordinates and label electron events according to their corresponding aperture positions, indicating the entrance positions of electron tracks.}
\label{fig:scan_mask}
\end{figure}

Our custom, focused-beam based alignment method for detector calibration at 200 keV enabled precise, sub-pixel localization of electron incident point. This was achieved by averaging the charge centroid positions of individual tracks to determine the corresponding electron entry locations (Figure \ref{fig:scan_mask}a), resulting in negligible bias and uncertainty compared to the pixel size. The electron flux was reduced accordingly to minimize multiple electron events per pixel and acquisition frame. Due to hundreds of scanning points, distributed randomly across the detector coordinates, an even distribution of impact positions within one sensor pixel was ensured. 

The second approach employed a mask with holes under parallel illumination conditions. Here, the grid of hole coordinates, accurately measured in the SEM, was fitted and used to label electron entrance positions accordingly. Even with a 25 $\mu$m thick tungsten foil, some signal bleed-through from the mask was still observed, which reduced the signal-to-noise ratio in the dataset. 
Therefore, for higher acceleration voltages, we recommend using a foil thickness of at least 30 – 35 $\mu$m. Nevertheless, the Mask 2 design performed well for generating training data and DQE curves at 60 keV and 80 keV (see Figure \ref{fig:scan_mask}b), while the Mask 3 design most effectively minimized overlap between individual electron tracks and was successfully employed at 120 keV. 

Overall, training datasets obtained using both methods demonstrated enhanced and unbiased spatial resolution. While the mask-based method permits acquisition of data in parallel mode, making it more time and data efficient, in the case of some HPDs one can also use region of interest-based readout, which can similarly reduce the amount of data required for the focused-beam approach. 
The number of scanning points and mask apertures was adjusted to ensure sufficient generalization capability of the resulting neural networks.

\subsection{\textbf{Pretreatment of experimental training data}}

For electron clusters associated with a given scan point or hole, we first identified the pixel, sharing the largest number of hits to obtain an approximate true position. Each electron cluster was then associated with that true position by requiring that it contained the selected pixel. The average charge-weighted centroid of the associated cluster was used to provide an unbiased and precise estimate of the ground-truth position, given sufficient statistics. Furthermore, we computed the root-mean-square error (RMSE) of the charge-weighted centroid distribution and excluded outlier clusters whose charge-weighted centroid deviated from the true position by more than three times the RMSE. Finally, pile-up clusters were removed by requiring that the cluster energy be lower than 1.1 times the nominal electron energy.

The algorithm described above was identical for both training dataset acquisition methods.

\subsection{\textbf{Comparison between the simulation and experimental training data}}\label{subsec1}

We observed discrepancies between synthetic and experimental training samples and examined these in details. Our analysis revealed that conventional simulation methods are inadequate for devices with a pixel pitch as small as that of the Mönch detector (i.e. 25 $\mu$m), likely because they ignore charge carrier dynamics, including electrostatic repulsion effects\cite{Xie2026OptimizationEffect}. 

Experimental data obtained via both methods yielded consistent detector response patterns and enabled efficient training of the neural network. The resulting sub-pixel reconstruction precision was comparable to, and in some cases exceeded, the performance achieved in our previous work \cite{Xie2024EnhancingLearning}. These results confirm that the proposed data acquisition strategies provide sufficiently accurate ground truth information for model training and calibrations. 

Furthermore, the experimental data facilitated optimization and extension of the simulation method to incorporate relevant physical processes serving as a validation benchmark for the comparison between simulation and experimental data. 

In this study, we concentrate exclusively on establishing and validating the data acquisition strategies necessary for generating experimental training datasets. A comprehensive quantitative evaluation—including modulation transfer function (MTF) and detective quantum efficiency (DQE) at various acceleration voltages will be presented separately, accompanied by electron microscopy measurements. 

\subsection{\textbf{Developed Machine Learning models and resolution enhancement}}

Deep learning based training and model implementation were performed using the PyTorch framework\cite{Paszke2019PyTorch:Library}. The model consisted of a backbone convolutional neural network (CNN) and a feature extraction model, described in details by Xie et al.\cite{Xie2024EnhancingLearning}.

For electron energies up to 120 keV, the estimated hole positions by averaging are biased toward the pixel center due to charge sharing in the sensor and as a consequence of limited lateral spread of signal due to the shorter trajectories of electrons within these lower energies.
To correct for this bias, we employed a grid-fitting procedure to accurately determine the corresponding positions of the aperture array on the detector plane.

For 200 keV electrons, models trained on synthetic data achieved a spatial resolution of 0.47 pixels, estimated using root mean square error (RMSE), representing 3.6-fold improvement over the charge centroid based method. The model trained on experimental data reached a spatial resolution of 0.6 pixels, corresponding to a 3-fold enhancement compared to the centroid approach. When the simulation-trained model was applied to predict impact positions from experimental data, a spatial resolution of 0.7 pixels was obtained. 

Overall, the experimentally trained model outperformed the simulation-based one when validated on experimental data, as expected. Thus, with our experimental training method, we achieved sub-pixel resolution of 0.6 pixels at 200 keV - a significant improvement over the conventional charge centroid technique with a spatial resolution of 1.8 pixels. 

\section{Conclusions}

In this work we developed two novel methods for experimental training data acquisition aimed at enhancing the spatial resolution of a hybrid pixel detector using machine learning. To our knowledge, this is the first study to enable experimental training data acquisition and to directly compare such datasets with synthetic training data.

While one of the proposed methods, employing a narrowly focused ($\simeq$2 $\mu$m) electron beam randomly scanned across the detector matrix, is specific to our probe-corrected microscope (i.e. NeoARM 200F), nonetheless being easily transferable to other microscopes (e.g. being potentially more straightforward with Titan series from ThermoFischer), the mask aperture based approach is more general and can be readily adapted to various microscope and detector configurations. By incorporating experimental training data, we achieved sub-pixel spatial resolution of 0.6 pixels for 200 keV electrons, representing a threefold improvement over the conventional charge centroid-based method. Accordingly, these developments pave the way toward universal, experimentally calibrated machine-learning frameworks for hybrid pixel detectors in both diffraction and imaging applications.
A detailed quantitative evaluation of spatial resolution, MTF/DQE characteristics, and imaging performance will be presented in a separate study integrating these calibration methods with electron microscopy data.

\newpage

\beginsupplement
\section{Appendix}
\vspace{-1em}

\begin{table}[!htbp]
\caption{Description of the mask geometry across the W-foils\label{tab1}}%
\begin{tabular*}{\columnwidth}{@{\extracolsep\fill}cccc@{\extracolsep\fill}}
\toprule
\textbf{Geometry}  & \textbf{Mask 1}  & \textbf{Mask 2} & \textbf{Mask 3}\\
\midrule
Diameter of holes    & 3 - 5 $\mu$m   & 5 $\mu$m  & 5 $\mu$m  \\
Distance between the holes    &  $>$ 200 $\mu$m    & 183 $\mu$m  & 283 $\mu$m  \\
Number of holes    & 12   & 400  & 576  \\
Thickness of the foil    & 20 $\mu$m  & 25 $\mu$m   & 25 $\mu$m   \\
\bottomrule
\end{tabular*}
\end{table}

\section{Simulation of shielding efficiency of W-foil}

\begin{figure}[htbp]
\centering
\includegraphics[width=0.5\textwidth]{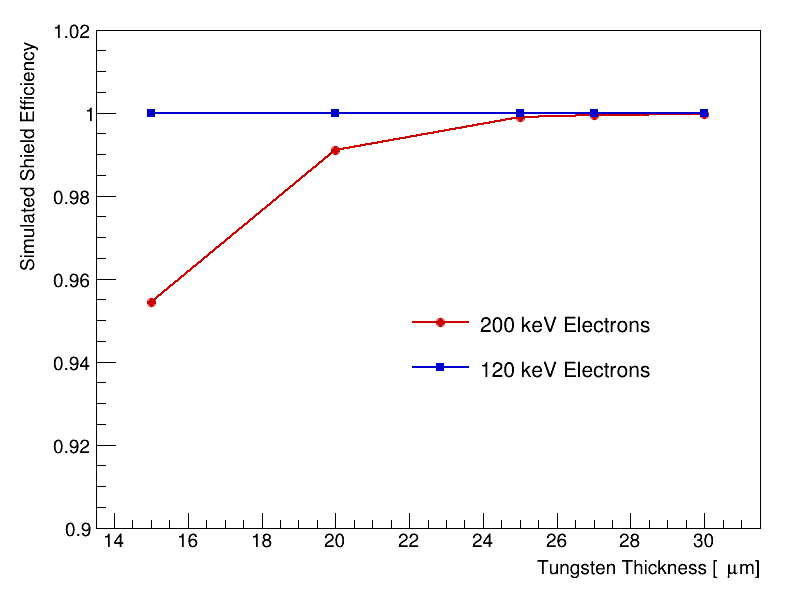}
\caption{\textbf{Shielding efficiency of W-foil at 120 and 200 keV.} Monte Carlo-based simulations of shielding efficiency at 120 keV (blue curve) and 200 keV (red curve), show that while the W-foil can be as thin as 15 $\mu$m for voltages $\leq$ 120 keV, for 200 keV, the optimal thickness is reached around 25 $\mu$m. However in practice we found thicknesses exceeding 25 $\mu$m to be more appropriate for voltages $\geq$200 keV.}
\label{fig:shielding}
\end{figure}

To evaluate the shielding efficiency of tungsten plates with varying thicknesses, simulations were performed using the Allpix$^2$ framework\cite{Spannagel2018Allpix2:Detectors}. A Mönch detector was modeled with the following design parameters: a pixel array of 400 × 400, a pixel pitch of 25 $\mu$m, and a 320 $\mu$m thick silicon sensor. A tungsten plate was positioned parallel to the detector surface at a distance of 5 mm, intercepting a 1 mm wide, normally incident electron beam. The shielding efficiency was quantified as 1$-$quantum efficiency of the simulated detector (see Figure \ref{fig:shielding}).

\section{Competing interests}
No competing interest is declared.

\section{Author contributions statement}
E.P. - writing original draft, conceptualization, methodology and experiments; X.X. - writing and editing, Monte-Carlo simulations, ML model design and validation, methodology and experiments; J.R. - writing and editing, mask fabrication with plasma FIB/SEM; L.B.F. - writing and editing, ML model design and validation; K.A.P. - writing and editing, methodology and experiments; B.B.H. - writing and editing, ML model design and validation; E.F. - writing and editing, methodology and experiments; A.B. - writing and editing; E.M. - writing and editing, conceptualization, methodology and experiments. E.M., E.P. and E.F. conceived the project and acquired funding.  

\section{Acknowledgments}
We thank Thomas King for his expertise in designing the mechanics to install Mönch detector on the microscope as well as for the installation of the aperture masks. We acknowledge the usage of the instrumentation provided by the Electron Microscopy Facility at PSI and we thank the EMF team for their help and support. We thank ScopeM team at ETH Zurich for the support and usage of instrumentation. The project was supported by the Swiss Data Science Center (SDSC) under the project number ML-ED [C21-02].

\bibliography{references}

@article{Pennycook2015,
    title = {{Efficient phase contrast imaging in STEM using a pixelated detector. Part 1: Experimental demonstration at atomic resolution}},
    year = {2015},
    journal = {Ultramicroscopy},
    author = {Pennycook, Timothy J. and Lupini, Andrew R. and Yang, Hao and Murfitt, Matthew F. and Jones, Lewys and Nellist, Peter D.},
    pages = {160--167},
    volume = {151},
    doi = {10.1016/j.ultramic.2014.09.013},
    issn = {18792723}
}

@article{Chen2021,
    title = {{Electron ptychography achieves atomic-resolution limits set by lattice vibrations}},
    year = {2021},
    journal = {Science},
    author = {Chen, Zhen and Jiang, Yi and Holtz, E. Megan and Odstr{\v{c}}il, Michal and Guizar-Sicairos, Manuel and Hanke, Isabelle and Ganschow, Steffen and Schlom, G. Darrell and Muller, A. David},
    number = {May},
    pages = {826--831},
    volume = {831}
}

@article{Pelz2017,
    title = {{Low-dose cryo electron ptychography via non-convex Bayesian optimization}},
    year = {2017},
    journal = {Scientific Reports},
    author = {Pelz, Philipp Michael and Qiu, Wen Xuan and B{\"{u}}cker, Robert and Kassier, Günther and Miller, R. J.Dwayne},
    number = {1},
    pages = {1--13},
    volume = {7},
    doi = {10.1038/s41598-017-07488-y},
    issn = {20452322},
    pmid = {28851880},
    arxivId = {1702.05732}
}

@article{Zhou2020,
    title = {{Low-dose phase retrieval of biological specimens using cryo-electron ptychography}},
    year = {2020},
    journal = {Nature Communications},
    author = {Zhou, Liqi and Song, Jingdong and Kim, Judy S. and Pei, Xudong and Huang, Chen and Boyce, Mark and Mendon{\c{c}}a, Luiza and Clare, Daniel and Siebert, Alistair and Allen, Christopher S. and Liberti, Emanuela and Stuart, David and Pan, Xiaoqing and Nellist, Peter D. and Zhang, Peijun and Kirkland, Angus I. and Wang, Peng},
    number = {1},
    pages = {1--9},
    volume = {11},
    publisher = {Springer US},
    url = {http://dx.doi.org/10.1038/s41467-020-16391-6},
    doi = {10.1038/s41467-020-16391-6},
    issn = {20411723},
    pmid = {32487987}
}

@inproceedings{Dinapoli2014,
    title = {{M{\"{O}}NCH, a small pitch, integrating hybrid pixel detector for X-ray applications}},
    year = {2014},
    booktitle = {Journal of Instrumentation},
    author = {Dinapoli, R. and Bergamaschi, A. and Cartier, S. and Greiffenberg, D. and Johnson, I. and Jungmann, J. H. and Mezza, D. and Mozzanica, A. and Schmitt, B. and Shi, X. and Tinti, G.},
    number = {5},
    volume = {9},
    doi = {10.1088/1748-0221/9/05/C05015},
    issn = {17480221}
}

@article{Sha2023,
    title = {{Ptychographic measurements of varying size and shape along zeolite channels}},
    year = {2023},
    journal = {Science Advances},
    author = {Sha, Haozhi and Cui, Jizhe and Li, Jialu and Zhang, Yuxuan and Yang, Wenfeng and Li, Yadong and Yu, Rong},
    number = {11},
    pages = {1--8},
    volume = {9},
    publisher = {American Association for the Advancement of Science},
    url = {https://doi.org/10.1126/sciadv.adf1151},
    doi = {10.1126/sciadv.adf1151}
}

@article{McMullan2023,
    title = {{Structure determination by cryoEM at 100 keV.}},
    year = {2023},
    journal = {Proceedings of the National Academy of Sciences of the United States of America},
    author = {McMullan, Greg and Naydenova, Katerina and Mihaylov, Daniel and Yamashita, Keitaro and Peet, Mathew J and Wilson, Hugh and Dickerson, Joshua L and Chen, Shaoxia and Cannone, Giuseppe and Lee, Yang and Hutchings, Katherine A and Gittins, Olivia and Sobhy, Mohamed A and Wells, Torquil and El-Gomati, Mohamed M and Dalby, Jason and Meffert, Matthias and Schulze-Briese, Clemens and Henderson, Richard and Russo, Christopher J},
    number = {49},
    pages = {e2312905120},
    volume = {120},
    url = {http://www.ncbi.nlm.nih.gov/pubmed/38011573},
    doi = {10.1073/pnas.2312905120},
    issn = {1091-6490},
    pmid = {38011573}
}

@article{Spannagel2018Allpix2:Detectors,
    title = {{Allpix2: A modular simulation framework for silicon detectors}},
    year = {2018},
    journal = {Nuclear Instruments and Methods in Physics Research, Section A: Accelerators, Spectrometers, Detectors and Associated Equipment},
    author = {Spannagel, S. and Wolters, K. and Hynds, D. and Alipour Tehrani, N. and Benoit, M. and Dannheim, D. and Gauvin, N. and N{\"{u}}rnberg, A. and Sch{\"{u}}tze, P. and Vicente, M.},
    month = {9},
    pages = {164--172},
    volume = {901},
    publisher = {Elsevier B.V.},
    doi = {10.1016/j.nima.2018.06.020},
    issn = {01689002},
    arxivId = {1806.05813}
}

@article{Jones2018AnPre-field,
    title = {{An optical configuration for fastidious STEM detector calibration and the effect of the objective-lens pre-field}},
    year = {2018},
    journal = {Journal of Microscopy},
    author = {Jones, L. and Varambhia, A. and Sawada, H. and Nellist, P. D.},
    number = {2},
    month = {5},
    pages = {176--187},
    volume = {270},
    publisher = {John Wiley and Sons Inc},
    doi = {10.1111/jmi.12672},
    issn = {13652818},
    pmid = {29315554}
}

@article{Xie2024EnhancingLearning,
    title = {{Enhancing spatial resolution in M{\"{O}}NCH for electron microscopy via deep learning}},
    year = {2024},
    journal = {Journal of Instrumentation},
    author = {Xie, X. and Barba Flores, L. and Bejar Haro, B. and Bergamaschi, A. and Fr{\"{o}}jdh, E. and M{\"{u}}ller, E. and Paton, K. and Poghosyan, E. and Remlinger, C.},
    number = {1},
    month = {1},
    volume = {19},
    publisher = {Institute of Physics},
    doi = {10.1088/1748-0221/19/01/C01020},
    issn = {17480221}
}

@article{Mahmoudi2025ExperimentalDiffraction,
    title = {{Experimental determination of partial charges with electron diffraction}},
    year = {2025},
    journal = {Nature},
    author = {Mahmoudi, Soheil and Gruene, Tim and Schr{\"{o}}der, Christian and Ferjaoui, Khalil D. and Fr{\"{o}}jdh, Erik and Mozzanica, Aldo and Takaba, Kiyofumi and Volkov, Anatoliy and Maisriml, Julian and Paunovi{\'{c}}, Vladimir and van Bokhoven, Jeroen A. and Keppler, Bernhard K.},
    number = {8079},
    month = {9},
    pages = {88--94},
    volume = {645},
    publisher = {Nature Research},
    doi = {10.1038/s41586-025-09405-0},
    issn = {14764687},
    pmid = {40836092}
}

@article{LeBeau2008ExperimentalMicroscopy,
    title = {{Experimental quantification of annular dark-field images in scanning transmission electron microscopy}},
    year = {2008},
    journal = {Ultramicroscopy},
    author = {LeBeau, James M. and Stemmer, Susanne},
    number = {12},
    month = {11},
    pages = {1653--1658},
    volume = {108},
    doi = {10.1016/j.ultramic.2008.07.001},
    issn = {03043991}
}

@article{Chan2024High-resolutionDetector,
    title = {{High-resolution single-particle imaging at 100–200 keV with the Gatan Alpine direct electron detector}},
    year = {2024},
    journal = {Journal of Structural Biology},
    author = {Chan, Lieza M. and Courteau, Brandon J. and Maker, Allison and Wu, Mengyu and Basanta, Benjamin and Mehmood, Hev and Bulkley, David and Joyce, David and Lee, Brian C. and Mick, Stephen and Czarnik, Cory and Gulati, Sahil and Lander, Gabriel C. and Verba, Kliment A.},
    number = {3},
    month = {9},
    volume = {216},
    publisher = {Academic Press Inc.},
    doi = {10.1016/j.jsb.2024.108108},
    issn = {10958657},
    pmid = {38944401}
}

@article{vanSchayck2023IntegrationWorkflow,
    title = {{Integration of an Event-driven Timepix3 Hybrid Pixel Detector into a Cryo-EM Workflow}},
    year = {2023},
    journal = {Microscopy and Microanalysis},
    author = {van Schayck, J. Paul and Zhang, Yue and Knoops, Kèvin and Peters, Peter J. and Ravelli, Raimond B.G.},
    number = {1},
    month = {2},
    pages = {352--363},
    volume = {29},
    publisher = {Oxford University Press},
    doi = {10.1093/micmic/ozac009},
    issn = {14358115}
}

@article{Xie2026OptimizationEffect,
    title = {{Optimization and validation of charge transport simulation for hybrid pixel detectors incorporating the repulsion effect}},
    year = {2026},
    journal = {Nuclear Instruments and Methods in Physics Research, Section A: Accelerators, Spectrometers, Detectors and Associated Equipment},
    author = {Xie, X. and Bergamaschi, A. and Br{\"{u}}ckner, M. and Carulla, M. and Dinapoli, R. and Ebner, S. and Ferjaoui, K. and Gautam, V. and Greiffenberg, D. and Hasanaj, S. and Heymes, J. and Hinger, V. and Kedych, V. and King, T. and Li, S. and Lopez-Cuenca, C. and Mazzoleni, A. and Mezza, D. and Moustakas, K. and Mozzanica, A. and M{\"{u}}ller, M. and Mulvey, J. and Paton, K. A. and Ruder, C. and Schmitt, B. and Sieberer, P. and Silletta, S. and Thattil, D. and Zhang, J. and Fr{\"{o}}jdh, E.},
    month = {1},
    volume = {1081},
    publisher = {Elsevier B.V.},
    doi = {10.1016/j.nima.2025.170894},
    issn = {01689002}
}

@techreport{JonesPracticalRecording,
    title = {{Practical Aspects of Quantitative and High-Fidelity STEM Data Recording}},
    author = {Jones, Lewys}
}

@article{Skoupy2025Ptychoscopy:Ptychography,
    title = {{Ptychoscopy: a user friendly experimental design tool for ptychography}},
    year = {2025},
    journal = {Scientific Reports},
    author = {Skoupy, Radim and M{\"{u}}ller, Elisabeth and Pennycook, Timothy J. and Guizar-Sicairos, Manuel and Fabbri, Emiliana and Poghosyan, Emiliya},
    number = {1},
    month = {12},
    volume = {15},
    publisher = {Nature Research},
    doi = {10.1038/s41598-025-09871-6},
    issn = {20452322}
}

@techreport{Paszke2019PyTorch:Library,
    title = {{PyTorch: An Imperative Style, High-Performance Deep Learning Library}},
    year = {2019},
    author = {Paszke, Adam and Gross, Sam and Massa, Francisco and Lerer, Adam and Bradbury Google, James and Chanan, Gregory and Killeen, Trevor and Lin, Zeming and Gimelshein, Natalia and Antiga, Luca and Desmaison, Alban and Xamla, Andreas Köpf and Yang, Edward and Devito, Zach and Raison Nabla, Martin and Tejani, Alykhan and Chilamkurthy, Sasank and Ai, Qure and Steiner, Benoit and Facebook, Lu Fang and Facebook, Junjie Bai and Chintala, Soumith}
}

@article{Gruene2018SchnelleElektronenbeugung,
    title = {{Schnelle Strukturaufkl{\"{a}}rung mikrokristalliner molekularer Verbindungen durch Elektronenbeugung}},
    year = {2018},
    journal = {Angewandte Chemie},
    author = {Gruene, Tim and Wennmacher, Julian T. C. and Zaubitzer, Christan and Holstein, Julian J. and Heidler, Jonas and Fecteau‐Lefebvre, Ariane and De Carlo, Sacha and M{\"{u}}ller, Elisabeth and Goldie, Kenneth N. and Regeni, Irene and Li, Teng and Santiso‐Quinones, Gustavo and Steinfeld, Gunther and Handschin, Stephan and van Genderen, Eric and van Bokhoven, Jeroen A. and Clever, Guido H. and Pantelic, Radosav},
    number = {50},
    month = {12},
    pages = {16551--16555},
    volume = {130},
    publisher = {Wiley},
    doi = {10.1002/ange.201811318},
    issn = {0044-8249}
}

@article{Pelz2023SolvingTomography,
    title = {{Solving complex nanostructures with ptychographic atomic electron tomography}},
    year = {2023},
    journal = {Nature Communications},
    author = {Pelz, Philipp M. and Griffin, Sinéad M. and Stonemeyer, Scott and Popple, Derek and DeVyldere, Hannah and Ercius, Peter and Zettl, Alex and Scott, Mary C. and Ophus, Colin},
    number = {1},
    month = {12},
    volume = {14},
    publisher = {Nature Research},
    doi = {10.1038/s41467-023-43634-z},
    issn = {20411723},
    pmid = {38036516},
    arxivId = {2206.08958}
}

\end{document}